\documentclass[aps,prb,twocolumn,groupedaddress,showpacs]{revtex4-1}
\usepackage{graphicx}
\usepackage{dcolumn}
\usepackage{bm}
\usepackage{color}
\usepackage{url}
\begin{document}

\title{Pseudopotentials for correlated electron systems}

\author{J. R. Trail}\email{jrt32@cam.ac.uk}
\affiliation{Theory of Condensed Matter Group, Cavendish Laboratory, J
  J Thomson Avenue, Cambridge CB3 0HE, United Kingdom}

\author{R. J. Needs}
\affiliation{Theory of Condensed Matter Group, Cavendish Laboratory, J
  J Thomson Avenue, Cambridge CB3 0HE, United Kingdom}

\date{\today}

\begin{abstract}

  A scheme is developed for creating pseudopotentials for use in
  correlated-electron calculations.  Pseudopotentials for the light
  elements H, Li, Be, B, C, N, O, and F, are reported, based on data
  from high-level quantum chemical calculations.  Results obtained with
  these correlated electron pseudopotentials (CEPPs) are compared with
  data for atomic energy levels and the dissociation energies,
  molecular geometries and zero-point vibrational energies of small
  molecules obtained from coupled cluster single double triple (CCSD(T)) 
  calculations with large basis sets.  The CEPPs give better results in
  correlated-electron calculations than Hartree-Fock-based
  pseudopotentials available in the literature.

\end{abstract}



\maketitle

\section{Introduction}

Pseudopotentials (also known as effective core potentials) are used to
replace the chemically inert core electrons within electronic
structure calculations.  The use of pseudopotentials reduces the
number of electrons that must be treated explicitly, and reduces the
size of the basis set required to represent the one-electron orbitals.
Pseudopotentials are particularly useful for heavy elements because
eliminating the numerous core electrons greatly increases the
efficiency of the calculations.  Pseudopotentials are also used for
light elements, even for hydrogen, because it can be advantageous to
work with smooth pseudo-orbitals.

The atomic pseudopotentials that we describe here can be used with a
range of many-body techniques, including quantum chemical methods.
However, our interest is mainly in quantum Monte Carlo (QMC)
calculations using the variational quantum Monte Carlo (VMC) and the
more accurate diffusion quantum Monte Carlo (DMC)
methods.
\cite{Ceperley-Alder_1980,Foulkes_QMC_review,Casino_reference,
      Lester_review_2012}
These approaches can provide accurate energies with a computational
cost that scales with the number of particles, $N$, as approximately
$N^{3}-N^{4}$, which is better than other correlated wave function
methods.  The scaling of the computational cost of QMC calculations
with atomic number $Z$ is, however, approximately
$Z^{5}-Z^{6.5}$.\cite{Ceperley_1986,Ma_2005_all_electron_atoms} Using
a pseudopotential reduces the effective value of $Z$ and makes QMC
calculations feasible for heavy atoms.  The relatively small memory
requirements of QMC calculations also facilitates applications on
massively parallel computers.

In many quantum chemical methods it is possible to use a frozen-core
approximation to reduce the computational effort, but it has not so
far proved possible to implement a frozen-core approximation in a
straightforward manner within the VMC and DMC methods, and therefore
pseudopotentials have been used instead.  In addition,
core-polarization effects can be included within the pseudopotential
formalism,\cite{muller_1984,shirley_1993,lee_2000} so that the
calculations go beyond the frozen-core approximation.  The plane wave
basis sets often used in Hartree-Fock (HF) and density functional
theory (DFT) calculations for extended systems have the advantage that
they can be improved systematically by increasing a single parameter,
the plane wave cutoff energy.  Correlated electron calculations have
recently been performed using plane wave basis
sets,\cite{Booth_2012_FCIQMC} for which pseudopotentials are almost
always used to ensure that the required basis set is of a 
reasonable size.

Calculations using smooth HF-based pseudopotentials
\cite{Trail_2005_asymptotic,Trail_2005_pseudopotentials,
      Burkatzki_2007,Burkatzki_2008,BFD_website}
have yielded good results in correlated electron calculations, which
appear to be superior to those obtained with DFT-based
pseudopotentials.\cite{Greeff_1998} The HF-based pseudopotentials do
not include correlation effects, and it is likely that more accurate
pseudopotentials could be developed for correlated-electron
calculations by starting from a correlated-electron theory.
Acioli and Ceperley\cite{acioli_1994} showed that pseudopotentials can
be formulated within many-body theory using the density matrix.  

In this paper we describe a scheme based on the density matrix for
generating atomic pseudopotentials using data from correlated electron
calculations.  Correlated electron pseudopotentials (CEPPs) are
generated for the H, Li, Be, B, C, N, O and F atoms.
We choose first row atoms as they allow the use of the same 
basis set for AE and pseudopotential calculations, resulting in 
a consistent basis set error.
For second row atoms and beyond such a choice is in principle 
possible, but in practice is numerically problematic.
Neither He or Ne are considered due to the absence of experimental 
molecular data.
To generate the CEPPs we use a combination of data from 
multi-configuration Hartree-Fock (MCHF)\cite{fischer_1997} atomic 
calculations performed on a radial grid using the \textsc{ATSP2K} 
code\cite{atsp2k}, and \emph{ab initio} data for atomic core 
polarizabilities.  The CEPPs are tested by comparing with results 
obtained from all-electron (AE) coupled-cluster single double triple
CCSD(T) calculations performed with large Gaussian basis sets and 
the MOLPRO\cite{molpro} code.
We perform tests for atomic energy levels, ionization energies and 
electron affinities, and for the relaxed geometries, well
depths, and total zero-point vibrational energies (ZPVEs) of 35
small molecules.
We also compare our CCSD(T) results with a set of
highly accurate experimental, semi-empirical, and \emph{ab initio}
data, which are referred to in the text as `accurate'.  The
semi-empirical results are used only when they have been considered to
be more accurate than the available experimental data.

We have constructed our CEPPs using atomic states
with a single valence electron, which are therefore ionized
states for all atoms considered, except H and Li. There is no need to 
use highly ionized states to generate pseudopotentials within single
particle theories, such as HF and DFT, because the construction of the 
pseudo-atom and resulting pseudopotential is straightforward.
In the many-body case, however, the construction of the pseudo-density
matrix and associated pseudo-Hamiltonian describing an atom with more 
than one valence electron is a challenging task, as discussed 
in Sec.\ \ref{theory_pseudo_atoms}.  The error due to transferring a
pseudopotential from the highly ionized environment in which it is
constructed to neutral or near neutral environments is partially
corrected by modifying the ionic calculations.
The remaining error is part of that tested by our CCSD(T) calculations.

The paper is arranged in three parts.  In Sec.\ \ref{CEPPs_theory} we
describe our method for generating the CEPPs, which is divided into
subsections on: \ref{theory_pseudo_atoms} theory of many-body
pseudo-atoms; \ref{single-valence-electrons} theory of
pseudopotentials for single-valence-electron atoms;
\ref{details_generation} a discussion of more subtle aspects of the
generation and implementation; and \ref{parameterization}
parameterization in a standard form for quantum chemistry packages.
The results obtained with various pseudopotentials are presented in
Sec.\ \ref{results}, which is divided into subsections on:
\ref{energy_levels} energy levels for isolated atoms with a
single-valence electron; \ref{ion_aff} ionization energies and
electron affinities for isolated atoms; \ref{geometry} optimum
geometries for small molecules; \ref{atomization} atomization energies
for small molecules; and \ref{vibrational} ZPVEs for small molecules.
Results for optimum geometries, atomization energies (with ZPVEs
removed), and ZPVEs for small molecules, are compared with AE CCSD(T)
data, `accurate' values, and results obtained using uncorrelated
pseudopotentials available in the literature.  The paper is concluded
by a short summary.
 
Atomic units are used, unless otherwise indicated.

\section{Method for generating the CEPPs \label{CEPPs_theory}}

\subsection{Theory of many-body pseudo-atoms \label{theory_pseudo_atoms}}

The many-body density matrix for a $p$-electron atom is
\begin{equation}
  \Gamma^p(\mathbf{r}_1\ldots\mathbf{r}_p ; \mathbf{r}_1'\ldots\mathbf{r}_p')=
  \Psi^*(\mathbf{r}_1\ldots\mathbf{r}_p) \, \Psi(\mathbf{r}_1'\ldots\mathbf{r}_p'),
\end{equation}
where $\Psi$ is a many-body wave function.  In order to reproduce the
scattering properties of the $p$-electron system using a
pseudopotential describing $n$ ($ \leq p$) ``valence'' electrons, we
define a spherical core region centered on the nucleus (that is, for
any $|\mathbf{r}_i| \leq r_c$).  We then reduce the full density
matrix to an $n$-electron density matrix outside of the core region
where all $|\mathbf{r}_i| > r_c$, and continue the $n$-electron
density matrix with some model form for any $|\mathbf{r}_i| \leq r_c$.
Details of this generalization of the one-body L\"uders
relation\cite{Luders_1955} to many-body systems are given by Acioli
and Ceperley\cite{acioli_1994}.

The reduction of the $p$-body density matrix to a $n$-body form is
accomplished by integrating over $p-n$ electron coordinates:
\begin{widetext}
\begin{equation}
\Gamma^n(\mathbf{r}_1\ldots\mathbf{r}_n ; \mathbf{r}_1'\ldots\mathbf{r}_n') =
  {p \choose n} \int d\mathbf{r}_{n+1}\ldots\mathbf{r}_p
  \Gamma^p( \mathbf{r}_1 \ldots\mathbf{r}_n ,\mathbf{r}_{n+1}\ldots\mathbf{r}_p ;
            \mathbf{r}_1'\ldots\mathbf{r}_n',\mathbf{r}_{n+1}\ldots\mathbf{r}_p ).
\end{equation}
\end{widetext}
$\Gamma^n$ is the $n$-body density matrix for $n$ electrons for all
$|\mathbf{r}_i| > r_c$, which is the quantity that must be preserved
so that the $n$-body pseudopotential reproduces the scattering
properties (to first order in the energy) of the $p$-body system.  To
reproduce the scattering of the all-electron system we define the
pseudo-density matrix outside of the core region by
\begin{eqnarray}
\tilde \Gamma^n(\mathbf{r}_1 \ldots \mathbf{r}_n ; \mathbf{r}_1'\ldots\mathbf{r}_n') =
       \Gamma^n(\mathbf{r}_1&\ldots&\mathbf{r}_n ; \mathbf{r}_1'\ldots\mathbf{r}_n') \nonumber \\*
  & &  \forall \ |\mathbf{r}_i|,|\mathbf{r}_i'| > r_c ,
\end{eqnarray}
with $\tilde \Gamma^n$ for any $ |\mathbf{r}_i|,|\mathbf{r}_i'| < r_c$
defined such that it is correctly normalized and well behaved.

Standard norm-conserving independent electron pseudopotentials
\cite{Hamann_pseudopotentials_1979} are defined using valence orbitals
only, but the reduced density matrix considered so far contains a
contribution from the core orbitals.  Consequently, for independent
electrons, a pseudopotential defined from the reduced density matrix
cannot be equivalent to the standard norm-conserving pseudopotential.
To allow the two to be equivalent we define a modified density matrix. 
We separate the underlying wave function into core and valence parts,
confine the core part to the core region, construct a $p$-body density
matrix, and reduce it to a modified $n$-body matrix.
This is achieved by expressing $\Psi$ as a sum of determinants
\begin{equation}
\Psi(\mathbf{r}_1\ldots\mathbf{r}_p) = \sum_{all} w_i D_i( \mathbf{r}_1\ldots\mathbf{r}_p ),
\end{equation}
where the $w_i$ are coefficients.  The determinants are formed from
the Natural Orbitals (NOs), $\psi_k$, which are eigenfunctions of the
density matrix with occupation numbers, $o_k$. The $n_c=(p-n)/2$ NOs 
with the largest occupation numbers, $o_k$, are identified as the core 
NOs.  
The determinants containing one or more core NOs are classified as
core determinants, $D^c_i$, while the others are classified as valence
determinants, $D^v_i$.  In this notation $\Psi$ is expressed as
\begin{eqnarray}
\Psi(\mathbf{r}_1\ldots\mathbf{r}_p) &=&
  \Psi_v( \mathbf{r}_1\ldots\mathbf{r}_p) + \Psi_c( \mathbf{r}_1\ldots\mathbf{r}_p) \\*
  &=&
  \sum_{j} w_j D^v_j( \mathbf{r}_1\ldots\mathbf{r}_p ) + 
  \sum_{j} w_j D^c_j( \mathbf{r}_1\ldots\mathbf{r}_p )      \nonumber
\end{eqnarray}
where $\Psi_v$ and $\Psi_c$ are taken to be the valence and core
components of the wave function.
Note that for the special case of a HF wave function the sole 
determinant would be classified as a core determinant.

In the next step the core NOs present in each $D^c_i$ are modified
such that they are zero outside of the core region, which gives a
modified wave function,
\begin{equation}
\Psi_m(\mathbf{r}_1\ldots\mathbf{r}_p) = \left\{ \begin{array}{ll}
\Psi_{v}(\mathbf{r}_1\ldots\mathbf{r}_p)              & \forall \ |\mathbf{r}_i| > r_c \nonumber \\
\Psi_{v}(\mathbf{r}_1\ldots\mathbf{r}_p) + 
\Psi_{c}(\mathbf{r}_1\ldots\mathbf{r}_p)              & \textrm{otherwise},
\end{array} \right.
\end{equation}
and a modified density matrix
\begin{equation}
\Gamma^p_m(\mathbf{r}_1\ldots\mathbf{r}_p ; \mathbf{r}_1'\ldots\mathbf{r}_p') =
   \Psi^*_{m}(\mathbf{r}_1\ldots\mathbf{r}_p) \Psi_{m}(\mathbf{r}_1'\ldots\mathbf{r}_p').
\end{equation}
This modification is desirable for the same reason as in the independent
particle case; the scattering properties of the pseudo-system are
modified by a very small amount, but significantly smaller values of
$r_c$ can be used. 
It is important to note that explicit modification of the core NOs is 
neither required nor used in the CEPP generation procedure described 
below, and the remaining non-core NOs are not modified.  This aspect of 
our scheme is essentially a many-body generalization of the removal of 
core orbitals in the generation of standard DFT or HF norm-conserving 
pseudopotentials.

Reduction to a $n$-body matrix with all coordinates constrained to lie
outside of the core region provides
\begin{widetext}
\begin{eqnarray}
\label{reduce}
\tilde \Gamma^n_m(\mathbf{r}_1\ldots\mathbf{r}_n ; \mathbf{r}_1'\ldots\mathbf{r}_n') &=&
  {p \choose n} \int d\mathbf{r}_{n+1}\ldots\mathbf{r}_p
  \Gamma^p_m( \mathbf{r}_1 \ldots\mathbf{r}_n ,\mathbf{r}_{n+1}\ldots\mathbf{r}_p ;
              \mathbf{r}_1'\ldots\mathbf{r}_n',\mathbf{r}_{n+1}\ldots\mathbf{r}_p ) \nonumber \\*
 &=&
  {p \choose n} \int d\mathbf{r}_{n+1}\ldots\mathbf{r}_p
     \Psi_{v}^*(\mathbf{r}_1 \ldots\mathbf{r}_n ,\mathbf{r}_{n+1}\ldots\mathbf{r}_p)
     \Psi_{v}  (\mathbf{r}_1'\ldots\mathbf{r}_n',\mathbf{r}_{n+1}\ldots\mathbf{r}_p)
  \ \forall \ |\mathbf{r}_i|,|\mathbf{r}_i'| > r_c .
\end{eqnarray}
\end{widetext}
An important property of this modification and reduction procedure is 
that although core determinants do not contribute to $\Gamma^p_m$ 
outside of the core region, they do contribute to $\tilde \Gamma^n_m$ 
outside of the core region.  It is this property that ensures that the 
CEPP is equivalent to a standard norm-conserving HF pseudopotential 
when the multi-determinant expansion is replaced by a single (core) 
determinant.  A proof of this property is given in the appendix.

The density matrix $\tilde \Gamma^n$ must be preserved for a 
$n$-body system to reproduce the scattering properties of the 
$p$-body all-electron atom, but $\tilde \Gamma^n_m$ must be preserved 
for a $n$-body system to reproduce the scattering properties of the 
$p$-body all-electron atom with the core electrons constrained to lie 
within the core region.

Note that the division of $\Psi$ into core and valence parts is 
not unique.  Our choice can, however, be justified by
noting that the NOs provide the most efficient orbital representation
of a wave function as a sum of determinants, and that the NO
occupation numbers associated with occupied (unoccupied) orbitals are
the largest (smallest) possible for all choices of
orbitals.\cite{davidson_1972}

In principle the next step should be to define the $n$-body density
matrix in the core region $|\mathbf{r}_i|,|\mathbf{r}_i'| < r_c$ in
such a manner that there are $n$ pseudo-electrons and the
density-matrix is smooth.  This is a highly non-trivial problem.  Even
if this could be solved, we would then need to invert the many-body
quantum Schr\"odinger equation to obtain a potential whose ground
state has density matrix $\tilde \Gamma^n_m$.  Again, this is a highly
non-trivial problem to which a solution would in general be a 
$n$-body effective potential.  To put this another way, in addition to
the familiar core-electron pseudo-interaction naturally identifiable
as a pseudopotential, there would also be core-electron-electron
pseudo-interactions, etc., up to and including a $n+1$-body
interaction.

To avoid this difficulty we consider only the special case of a single
valence electron.  The $n$-body modified density matrix is then given 
by
\begin{equation}
\label{onedm}
  \Gamma_m^1(\mathbf{r};\mathbf{r}') = \sum_{i>n_c} o_i\psi^*_i(\mathbf{r}) \psi_i(\mathbf{r}') 
  \ \ \textrm{for} \ |\mathbf{r}|,|\mathbf{r}'| > r_c ,
\end{equation}
where the $\psi_i$ are the non-core unmodified NOs with indices 
ordered by decreasing $o_i$.  Note that although a valid density 
matrix must be representable by a complete set of orthonormal 
functions, this condition does not invalidate Eq.\ (\ref{onedm}).
Orthonormality is allowed since our definition is over a subspace 
only, and completeness is allowed since we may introduce a subset 
of additional NOs with zero occupancy that do not contribute to the 
sum.

Since there is only one valence electron, the charge density
contains the same information as the density matrix and we can work 
in terms of the pseudo-density, $\tilde \rho$, defined as
\begin{equation}
\label{dens_def}
  \tilde \rho(\mathbf{r}) = \left\{ \begin{array}{ll}
  \Gamma_m^1(\mathbf{r};\mathbf{r})          & |\mathbf{r}| >    r_c \nonumber \\
  \phi^2(\mathbf{r})                         & |\mathbf{r}| \leq r_c ,
\end{array} \right.
\end{equation}
with $\phi^2$ constructed to give the required particle number and
satisfy various smoothness conditions.  Finding the one-body potential
for which $\tilde \rho$ is the ground state is then straightforward.

Before describing the implementation of such an inversion procedure, 
it is worth clarifying the separation of core and valence electrons 
used above.  This separation procedure is a natural generalization of 
mean-field norm-conserving pseudopotential theory to many-body theory, 
is equivalent for non-interacting electrons in a mean-field, 
and is no more \emph{ad hoc} than the well established procedure of 
removing core orbitals in mean-field pseudopotential theory.  The 
pseudopotential provided by the inversion procedure describes the 
effect of core electrons on valence electrons, including correlation.
From this perspective it is apparent that the CEPP fits naturally into 
the methodology of correlated wave function methods.  The effects of 
correlation involving core electrons is described by the external 
potential part of the (pseudo-)Hamiltonian, whereas correlation 
between valence electrons remains a direct consequence of 
electron-electron interactions.  For example, a `HF' calculation 
using CEPPs includes all correlation except that between valence 
electrons, and post-HF methods naturally add the correlation between 
valence electrons to provide a complete description of the correlation 
effects.

\subsection{Pseudopotentials for single-valence-electron
  systems \label{single-valence-electrons}}

To create a pseudopotential for a single-valence-electron system we
calculate a one-body potential that generates the density (and
therefore the one-body density matrix) of the all-electron correlated
atom outside of the core region.  We calculate the NOs and occupation
numbers on a numerical grid using the \textsc{ATSP2K}
code\cite{atsp2k}.  Core electron excitations are included in the
active space (AS) so as to describe core-valence and intra-core
correlation.  The finite AS manifests itself as the requirement that
only a finite number of orbitals can be included in the calculation,
with each indexed by an angular momentum eigenvalue $l$ and an index
analogous to the primary quantum number of hydrogenic atoms, $n$.  We
have employed the largest ranges of $n$ and $l$ that avoid numerical
errors arising from linear dependence problems which, in practice,
corresponds to $n \le 6$ and $l \le 3$, with the removal of one or two
$n=6$, high $l$ orbitals for Li, Be, and B.  These ranges, together
with the single and double excitations of the core and valence
electrons, define the AS used.

The lowest energy states of different symmetries are considered in
order to construct potentials for the $s$, $p$ and $d$ angular
momentum channels of a non-local pseudopotential.  For example, the
MCHF data for carbon are generated for the C\textsuperscript{3+} ion
in the $1s^22s\ (2S)$, $1s^22p\ (2P)$ and $1s^23d\ (2D)$
configurations.

We divide the radial space into three regions, in a manner similar to
Lee \emph{et al.}\cite{lee_2000}, defining the pseudopotential and
pseudo-orbital separately in each region.  The potential for each
angular momentum channel $l$ is written as
\begin{equation}
\label{eq:V_l}
V_l(r) = \left\{ \begin{array}{rl}
V_{I}(r)     & 0   \leq r < r_c  \\
V_{II}(r)    & r_c \leq r < r_0  \\
V_{III}(r)   & r_0 \leq r .
\end{array} \right.
\end{equation}
The core region is denoted by $I$, and the outer part is divided into
two regions, $II$ and $III$.  In principle the latter division is not
necessary, but it is useful in controlling the numerical problems that
arise far from the nucleus where the density decreases exponentially.

In region $III$ we take the potential to have the form 
\begin{equation}
\label{V_{III}}
V_{III} = -\frac{ Z_v }{r} -\frac{1}{2}\frac{\alpha}{r^4},
\end{equation}
which is the sum of terms due to the valence charge $Z_v=Z-2n_c$, and
core-valence correlation effects, with the latter described by a core
polarization potential (CPP).  The parameter $\alpha$ is the dipole
core-polarization parameter.  Higher order terms (corresponding to
quadrupole polarizability and higher) are not considered here since
they are negligible, provided $r_0$ is sufficiently large.  This form
provides the correct asymptotic potential far from the nucleus.

The pseudopotential in region $II$ is constructed directly from the
MCHF charge density.  A pseudo-orbital is defined by 
$u=\left[ \tilde \rho_l \right]^{1/2}$ 
and the pseudopotential is obtained by inverting the radial Schr\"odinger equation:
\begin{equation}
\label{eq:epsilon1}
V_{II} = \frac{1}{2} \frac{1}{ u } \frac{ d^2 u }{ dr^2 } - \frac{1}{2}\frac{ l(l+1) }{ r^2 }  + \epsilon.
\end{equation}
It is tempting to identify $\epsilon$ with some target eigenvalue, but
this is not necessary since the potential in region $I$ has not yet
been defined.  Instead we note that the potentials should be
continuous at $r=r_0$, which leads to
\begin{equation}
\label{eq:epsilon2}
\epsilon =
- \frac{ Z_v }{r_0} 
- \frac{1}{2}\frac{\alpha}{r_0^4} 
- \frac{1}{2} \left. \frac{1}{ u } \frac{ d^2 u }{ dr^2 } \right|_{r_0} 
+ \frac{1}{2}\frac{ l(l+1) }{ r_0^2 }.
\end{equation}
The potential has a very small discontinuity in its first derivative
at $r=r_0$.  The potential in region $II$+$III$ is thus defined as
that which reproduces the target charge density outside of the core
region.  Note that $r_0$ must be large enough for
$V_{III}$ of Eq.\ (\ref{V_{III}}) to be accurate, but small enough for
the inversion procedure to be numerically stable.
We used values of $r_0$ ranging from 37$r_c$ for H to 2.3$r_c$ for F.
The pseudo-orbital in $II$+$III$ is then generated as the solution of 
\begin{equation}
\label{eq:epsilon3}
\left[
-\frac{1}{2} \frac{ d^2 }{ dr^2 } + \frac{1}{2}\frac{ l(l+1) }{ r^2 } + V_{I} - \epsilon
\right] \phi(r) = 0,
\end{equation}
which is obtained by integrating from a large $r$ of about 50-100 
a.u.\ down to $r_c$.  At large $r$ the orbital is taken to be a 
decaying Whittaker function of the second kind, which is an exact 
solution of the Schr\"odinger equation with potential $-Z_v/r$.  
$\phi$ is then normalized in region $II$+$III$ such that
\begin{equation}
\label{eq:norm_conservation}
  4 \pi \int_{r_c}^{+\infty} \phi^2 r^2 dr = 4 \pi \int_{r_c}^{+\infty} u^2 r^2 dr.
\end{equation}

The pseudo-orbital in region $I$ is chosen to be of a standard
form\cite{Troullier_1991}
\begin{equation}
  \phi = r^{l+1} \exp \left[ \sum_{k=0}^6 a_{2k} \, r^{2k} \right] \ \ \ \ 0 \leq r < r_c,
\end{equation}
where the parameters $\{a_{2k}\}$ are defined by seven constraints:
continuity of the value and first four derivatives of $\phi$ at $r_c$,
$V''_I(0)=0$, and norm conservation in region $I$ associated with Eq.\
(\ref{eq:norm_conservation}),
\begin{equation}
 4 \pi \int_{0}^{r_c} \phi^2 r^2 dr = 1 - 4 \pi \int_{r_c}^{+\infty} u^2 r^2 dr.
\end{equation}
Inverting the Schr\"odinger equation in region $I$ provides the
potential as
\begin{equation}
\label{eq:epsilon4}
  V_{I} = \frac{1}{2} \frac{1}{ \phi } \frac{ d^2 \phi }{ dr^2 } 
  - \frac{1}{2}\frac{ l(l+1) }{ r^2 } + \epsilon.
\end{equation}
We now have the effective potential in regions $I$, $II$, and $III$
for which the lowest one-electron eigenstate with $\epsilon$ and
$\phi$ has the same norm as $u$ in region $I$.

In our scheme we provide a target density, $\tilde \rho$, in regions
$II+III$, which results in the density $\phi^2$ in regions $I+II+III$.
Before commencing we clarify the relationship between $\tilde \rho$
and $\phi^2$ in region $II+III$.
For the ideal case where an exact $\tilde \rho$ is available we
could take the limit $r_0 \rightarrow \infty$, so that $\alpha$ would
not be required (core polarization effects would still be present
since the AE wave function is correlated). 
Furthermore, we would have $\epsilon=-I$, where $I$ is the exact
ionization energy of the ion (for example, for carbon this would be the
difference between the total energies of C\textsuperscript{3+} and
C\textsuperscript{4+}), so we would exactly reproduce the ionization 
energy and the charge density outside of the core region.
For a limited AS (for example a HF AE calculation) the same desirable 
properties occur, as long as $r_0 \rightarrow \infty$ can be used
to eliminate region $III$.

Region $III$ exists when $r_0$ is finite, and the potential within
this region is approximate.
Consequently, $\epsilon \neq -I$, but it is an accurate approximation
since $\epsilon$ approaches $-I$ with increasing $r_0$, which ensures
that $\phi^2$ approaches $\tilde \rho$ with increasing $r_0$.  To put
this another way, provided that Eq.\ (\ref{V_{III}}) is a good
approximation, the pseudopotential will reproduce the AE ionization
energy and (modified) density matrix outside of the core region.

Note that the quantities that must be supplied for generating the
pseudopotential are the target density in region $II+III$, and the
core-polarization parameter, $\alpha$. A target eigenvalue is not
required.

\subsection{Fixed cores and the role of the CPP
\label{details_generation}}

In the above discussion the pseudopotential describes the frozen core
of an ion (e.g., C\textsuperscript{3+}), whereas it would be better
for the core implicit in the pseudopotential to be that of the neutral
atom, as this is closer to the electronic states of interest. This
difference is expected to be small, but we account for it by fixing
the core orbitals to be those of a correlated neutral atom.  The
required orbitals and energies are obtained from two different MCHF
calculations with the same AS, with all determinant expansion
coefficients free to vary.  First `relaxed core' MCHF results are
generated for the neutral atom, with all orbitals free to vary.  From
the resulting orbitals we select the $n_c$ core orbitals.
Next, we perform a `fixed core' MCHF calculation, including the 
core orbitals of the neutral atom in the AS, allowing no 
variation in the core orbitals, but with all the other orbitals free 
to vary.
This provides a new set of NOs, $\{\psi_i \}_{fxd}$, 
from which we construct the target density of Eq.\ (\ref{dens_def}), 
and hence the `fixed core' pseudopotential.  This
procedure is used for all of the CEPPs generated in this work, and it
provides pseudopotentials that include core-valence (and intra-core)
correlation effects.

Lee \emph{et al.}\cite{lee_2000} have generated pseudopotentials from
HF orbitals and `accurate' ionization energies of ions.  They evaluate
corrections to both the norm and ionization energy due to fixing the
core to be that of a neutral atom.  These HF corrections to the target
norm and eigenvalue are then used with the HF orbitals of the ion to
generate pseudopotentials.  The approach of Lee \emph{et
  al.}\cite{lee_2000} may be generalized from HF to MCHF, resulting in
a scheme similar to that given in this paper.
At first this approach seems appealing as it includes a semi-empirical
description of physical effects not present in the MCHF Hamiltonian
via the target ionization energies, such as relativistic and finite
nuclear mass effects.
However, we have not considered this option further since the 
performance of such pseudopotentials was similar or marginally worse 
than our more \emph{ab initio} pseudopotentials for all systems 
considered.

Some further steps are required to define the CEPP.  The
angular-momentum-dependent pseudopotential, $V_l$, (Eq.\ (\ref{eq:V_l}))
provides an \emph{ab initio} representation of the interaction between
a single valence electron and the core.  Although this could be
applied as it stands to systems containing many electrons and atoms,
consideration of the semi-empirical theory of CPPs strongly suggests
that we can improve on this.

So far we have used the term `core-valence correlation' to refer to
all atomic correlation not represented by a HF pseudopotential.
Consideration of semi-empirical CPPs allows us to separate this
correlation into different parts, and to include the correlation
(mostly due to exchange) arising from interactions between the
separate cores of a many-atom system.  The potential part of the
Hamiltonian, including the CPP, can be written as
\begin{equation}
\label{hamilcpp}
\hat V_{cv} = \left[ \hat V_v + \hat V_{e} \right] + \left[ \hat V_{e-e} + \hat V_{e-I} + \hat V_{I-I} \right],
\end{equation}
where $\hat V_{cv}$ represents core-core, core-valence and
valence-valence correlation effects, $\hat V_v$ represents 
valence-valence correlation effects, and the four CPP terms 
provide an approximate representation of core-core and 
core-valence correlation effects as an induced dipole interaction.
Considering the CPP terms separately, 
$\hat V_{e}$ describes a one-body valence electron potential,
$\hat V_{e-e}$ describes a two-body valence electron potential,
$\hat V_{e-I}$ describes a three-body core-core-electron interaction,
and $\hat V_{I-I}$ describes a two-body inter-core interaction.
Generating a pseudopotential from an isolated atom can directly
provide \emph{ab initio} versions of $\hat V_{e}$ and $\hat V_{e-e}$,
and, since $V_l$ is a one-body potential generated from an ion with a
single valence electron, it only provides an \emph{ab initio} version 
of
\begin{equation}
\hat V_{cv} = \left[ \hat V_v + \hat V_{e} \right].
\end{equation}
The $V_l$ pseudopotential version of this quantity can be expected to 
be superior to the semi-empirical CPP description.

In light of this we use $V_l$ to provide an \emph{ab initio} description 
of the one-body valence electron potential, but rather than ignore the 
other parts of the correlation we describe them using part of the 
semi-empirical CPP interaction.
This is achieved by subtracting the one-body core-electron part of the
semi-empirical CPP interaction from the \emph{ab initio}
pseudopotential $V_l$, and adding it back in as part of the full CPP
correction.

M\"uller and Meyer\cite{muller_1984} describe alkali dimers using
configuration interaction calculations together with a CPP, and report
the influence of each of the CPP terms on the equilibrium geometry and
binding energy.
They find the $\hat V_{e}$ and $\hat V_{e-I}$ terms to be the largest and 
of opposite sign.
These results suggest that it is desirable to include a semi-empirical 
representation of the correlation that is not included in $V_l$.

We use the parameterized CPPs of Shirley and Martin\cite{shirley_1993}, 
and define a modified pseudopotential
\begin{equation}
\label{removecpp}
V^{pp}_l = V_l - \hat{V}_{e}                       
         = V_l + \frac{1}{2} f^2(r/\bar r_l) \frac{\alpha}{r^4},
\end{equation}
where $f(x)=(1-e^{-x^2})^2$ is a short range truncation function used
to remove the unphysical singularity at $r=0$, 
the $\bar r_l$ are empirically defined cutoff radii for this truncation 
function,
and $\alpha$ is the polarizability parameter used in generating $V_l$.
Throughout this paper the acronym CEPP refers to $V^{pp}_l$ used
together with the CPP, and we emphasise that the one-body
valence-electron potential part of the CEPP is \emph{ab initio} even
though the one-body valence electron potential part of the CPP is
semi-empirical.

\subsection{Pseudopotential parameterization \label{parameterization}}

We tabulate each pseudopotential on a radial grid for use in QMC
applications.
However, most quantum chemistry packages require a Gaussian
parameterization of the pseudopotential.  
Following the same procedure as Trail and Needs
\cite{Trail_2005_asymptotic,Trail_2005_pseudopotentials}, we use the
parameterization
\begin{equation}
\label{pp_param}
\tilde{V}^{pp}_l=
\sum_{q=1}^{6} A_{ql} r^{n_{ql}} e^{-a_{ql}r^2} = \left\{ \begin{array}{rl}
Z_v/r + V^{pp}_{local}           & l =    local  \\
V^{pp}_l - V^{pp}_{local}        & l \neq local ,
\end{array} \right.
\end{equation}
where $n_{ql}=0$, except for the local channel, for which $n_{ql}=-1$
for $q=1$, $n_{ql}=+1$ for $q=6$, and $n_{ql}=0$ otherwise.  The local
channel is chosen to be $l=2$.

The values of the parameters $A_{ql}$ and $a_{ql}$ are obtained by
optimization.  The number of parameters is reduced by imposing the
constraints
\begin{equation}
\tilde{V}^{pp}_l(0) - V^{pp}_l(0) =
\tilde{V'}^{pp}_l(0) =
\tilde{V''}^{pp}_l(0) = 0 ,
\end{equation}
which ensures that the parameterized form is approximately equal to
the tabulated potential close to $r=0$.  These constraints fix four
and two $A_{ql}$ parameters, respectively, in the local and non-local
channels.

The initial set of parameter values is obtained by taking 100 random
values distributed uniformly over a suitable range, performing a least
squares fit for each set, and solving for the ground state of each
fitted potential.  The set of parameters with the lowest error in the
eigenvalue are then chosen, and used as initial values for the
minimization of the penalty function\cite{barthelat_1977}
\begin{equation}
\Sigma_1=\langle \phi_l | \left[
\tilde \epsilon_l | \tilde \phi_l \rangle \langle \tilde \phi_l | -
       \epsilon_l |        \phi_l \rangle \langle        \phi_l |
\right]^2 | \phi_l \rangle ,
\end{equation}
where $(\phi_l,\epsilon_l)$ and $(\tilde \phi_l,\tilde \epsilon_l)$
are the lowest energy eigenstates for the tabulated and parameterized
pseudopotential, respectively.  Minimization of $\Sigma_1$ is
terminated when $1- \langle \tilde \phi_l | \phi_l \rangle < 10^{-6}$
and $ | \tilde \epsilon_l - \epsilon_l | < 10^{-5}$ a.u.  
A second penalty function,
\begin{equation}
\Sigma_2=\sum_{i=1}^5 ( \tilde \epsilon_{li} - \epsilon_{li} )^2,
\end{equation}
was then minimized until $\max_i | \tilde \epsilon_{li} -
\epsilon_{li} | < 10^{-8}$ a.u.  Only the lowest five states are
included in $\Sigma_2$, since higher states are essentially hydrogenic
and their energies are largely independent of the quality of the
pseudopotential.  The initial least-squares fitting, choice of best
fit, and minimization of $\Sigma_1$ are employed as an effective way
of selecting a good minimum among many.  The second optimization stage
is used to improve a useful measure of the quality of the
parameterization, $\Sigma_2$, without significantly increasing
$\Sigma_1$.

\section{Results \label{results}}

We generate CEPPs for H, Li, Be, B, C, N, O, and F.

For Li, Be, B, C, N, O, and F we choose there to be two core 
electrons and use the core-polarization parameter values of Shirley 
and Martin\cite{shirley_1993}.
The three pseudopotential channels are generated from the densities
provided by MCHF calculations for the $1s^22s\ (^2S)$, $1s^22p\
(^2P)$, and $1s^23d\ (^2D)$ configurations, so that the CEPPs are
generated from the neutral Li atom, and the Be\textsuperscript{1+},
B\textsuperscript{2+}, C\textsuperscript{3+}, N\textsuperscript{4+},
O\textsuperscript{5+}, and F\textsuperscript{6+} ions.  
The core radii are taken from our earlier Dirac-Fock (DF)
pseudopotentials\cite{Trail_2005_pseudopotentials}.

\begin{figure}[t]
\includegraphics[scale=1.00]{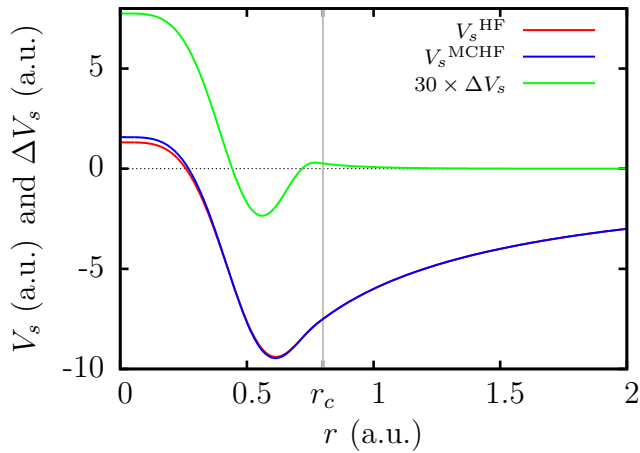}
\caption{ \label{fig1} CEPPs (before removal of the CPP) constructed
  from the O$^{5+}$ state.  The pseudopotentials are generated from
  MCHF and HF densities, and the MCHF pseudopotential minus the HF
  pseudopotential (with identical core radii and other parameters), is
  shown scaled by $\times 30$. }
\end{figure}

Figure~\ref{fig1} shows CEPPs constructed using both the MCHF and HF 
charge densities of O$^{5+}$ with a single $s$ electron in the lowest 
energy level.
The differences between the two potentials are very small on the scale
of the depth of the pseudopotential.
The differences near $r=0$ are unimportant, but the deeper potential 
around $r=0.6$ a.u.\ of the MCHF-based CEPP is significant.
The sixth ionization energy of the MCHF-based CEPP is 
$0.074$ eV larger than that of the HF-based CEPP, and the
norm (fraction of the orbital within the pseudopotential core radius
of $0.8$ a.u.) is about $0.00033$ larger for the MCHF-based CEPP.
The MCHF-based CEPP is therefore slightly more attractive than the
HF-based CEPP for this electronic state.

For H we choose there to be no core electrons, and hence $\alpha=0$.
The three pseudopotential channels are generated from the MCHF
densities (the MCHF and HF Hamiltonians are identical for the H atom)
for the neutral $1s\ (^2S)$, $2p\ (^2P)$, and $3d\ (^2D)$
configurations.  The core radii, $r_c$, are taken to have the same
values as in our earlier work\cite{Trail_2005_pseudopotentials}, with
the exception of the H $d$-channel, for which we use a core radius of
$r_c=0.8$ a.u.

In this section we assess the accuracy with which these
pseudopotentials can reproduce AE and `accurate' data.  We begin by
evaluating energy levels for single-valence-electron atoms and ions,
in order to assess the quality with which the CEPPs describe a simple
system without valence-valence or inter-core correlation.  These
calculations are performed by direct numerical integration of the
radial equation.  Next we move on to atoms with many valence
electrons, obtaining results using the MOLPRO\cite{molpro}
implementation of CCSD(T) theory (and RCCSD(T) when required) with
large Gaussian basis sets.  We calculate atomic ionization energies
and electron affinities at this level of theory, in order to assess
the quality of the CEPPs for isolated atoms.  Finally, we perform
CCSD(T) (and RCCSD(T)) calculations on a test set of 35 small
molecules, chosen by taking neutral members of the G1\cite{G1} set,
removing those containing atoms other than H, Li, Be, B, C, N, O, 
and F, and adding H$_2$, BH, Be$_2$, B$_2$, C$_2$, and NO$_2$.  For 
each molecule we evaluate: ($i$) the optimum geometry by energy
minimization; ($ii$) the well depth, defined as the sum of the 
atomization energy and zero-point vibrational energy and denoted $D_e$, 
and ($iii$) the ZPVEs.

The CCSD(T) calculations are performed using Dunning basis
sets\cite{basis}.
Basis set contraction is not employed in order to allow the AS for 
the AE and pseudopotential calculations to be as consistent as 
possible, but also flexible enough to provide high accuracy.
The AS for the AE calculations is defined to include core excitations. 
The computational cost of achieving a given level of accuracy could be 
reduced significantly by generating a smaller contracted even-tempered 
basis set for each CEPP using the Dunning model.  Although the 
computational efficiency provided by such a choice would be desirable 
in future applications of the CEPPs, it does not provide the 
consistent error appropriate for assessing the accuracy with which the 
CEPP Hamiltonian reproduces AE results.

Estimates of the complete basis set limits of the total energies were
obtained by extrapolating the $n=3,4$ basis set energies (for example,
cc-pwCVTZ and cc-pwCVQZ) using the non-rigorous formulae
\begin{eqnarray}
\label{extrap}
E_1(n) &=& a_1 + b_1 e^{-n}   \nonumber \\
E_2(n) &=& a_2 + b_2 n^{-3}   \nonumber \\
E_{est} &=& (a_1+a_2)/2 \pm (a_1-a_2)/2,
\end{eqnarray}
where $E_1$ and $E_2$ underestimate and overestimate the correlation
energy, respectively, and half their sum and half their difference
provide estimates of the converged energy and error bounds.  
This is essentially a simplified version of the approach of Feller
\emph{et al.}\cite{Feller_2010}.  Extrapolation is not used in
calculating geometries or ZPVEs.

The ionization energies and electron affinities are calculated using
the uncontracted aug-cc-pVnZ basis set for H, and the uncontracted
aug-cc-pwCVnZ basis sets for other atoms, with $n=T,Q$.
For the molecules we use uncontracted basis sets:
aug-cc-pVnZ for H$_2$, 
aug-cc-pwCVnZ and aug-cc-pVnZ for LiH, 
aug-cc-pwCVnZ for Li$_2$, and 
cc-pwCVnZ for all others.
We usually take $n=Q$ for geometry optimization and calculating ZPVEs,
and $n=T,Q$ for total energies.
Exceptions are described as they arise, and the well depths
are evaluated using isolated neutral atom calculations 
performed with the same basis as for the molecule.

We compare the performance of the CEPPs within CCSD(T) with AE
results, and results obtained using two other types of
pseudopotential, the norm-conserving DF pseudopotentials of Trail and
Needs\cite{Trail_2005_pseudopotentials} (TNDF), and the
scalar-relativistic energy-consistent HF pseudopotentials of Burkatzki
\emph{et al.}\cite{Burkatzki_2007} (BFD).  Both the BFD and TNDF
pseudopotentials include relativistic effects, while the CEPPs do not.
This comparison is used to
investigate the changes that arise from the explicit inclusion of
correlation in the pseudopotential generation process, and to test
whether the CEPPs generated from single-electron ions transfer
successfully to systems with multiple valence electrons and atoms.
Burkatzki \emph{et al.}\ provide contracted basis sets for use with
their pseudopotentials.  We do not use these basis sets, since the
uncontracted Dunning basis sets we have used consistently provide
better convergence properties and lower energies for first row
diatomic and hydride molecules.

We take our baseline data to be that provided by AE CCSD(T)
calculations, defining the error in the well depth, $D_e$, the
geometry, and the ZPVEs as the deviation of the pseudopotential
results from the AE ones.  However, in each plot the `accurate' data
are also shown as deviations from the baseline and are discussed
separately at the end of this section.  We make our primary comparison
of pseudopotential results with the AE baseline data in order to
separate the pseudopotential errors from those present in both the AE
and pseudopotential CCSD(T) calculations (for example, errors due to
the finite basis sets, lack of relativistic effects, use of harmonic
vibrational energies, etc).  This is desirable since the magnitudes of
the errors due to using the CEPPs and the CCSD(T) theory are of
comparable size.

Before testing the pseudopotentials, we briefly
mention the accuracy of the MCHF calculations used to generate the
pseudopotentials, and the RCCSD(T) results for the same AE atoms and
ions.
We quantify the accuracy by expressing the difference between the 
calculated and HF total energies as a fraction of the difference 
between the `accurate' and HF total energies.
For the three-electron atoms and ions used to construct the
pseudopotentials for Li, Be, B, C, N, O, and F, MCHF provides
$\sim$98--100\ \% of the correlation energy, whereas RCCSD(T) provides
$\sim$100\ \% for all of the atoms and ions.
For the neutral atoms MCHF provides $\sim$98--88\ \% of the correlation, 
decreasing monotonically with increasing atomic number, and 
RCCSD(T) provides $\sim$100\ \% of the correlation energy for all 
of the atoms.

\subsection{Energy levels for single-valence-electron atoms and ions
\label{energy_levels}}

The accuracy of the CEPPs is first tested by calculating
single-electron excitation energies using numerical integration.
These energies are compared with `accurate' energy levels by 
averaging spectral data\cite{NIST_De} over the spin-splitting using
the standard formula
\begin{equation}
\label{spnavg}
\epsilon_{nl}=\left[ (l+1)\epsilon_{j=l+1/2} + l\epsilon_{j=l-1/2} \right] / \left[ 2l+1 \right],
\end{equation}
and subtracting the $n\rightarrow \infty$ limit (taken from
NIST\cite{NIST_De}).  Equation\ (\ref{spnavg}) is a natural choice as
it would exactly remove spin-splitting if it arose from a perturbative
treatment of spin-orbit coupling.  
The difference between the energy levels obtained with the tabulated
CEPP and the `accurate' energy levels provides a measure of the
accuracy with which a pseudopotential reproduces the one-electron
excitation spectrum.

\begin{figure}[t]
\includegraphics[scale=1.00]{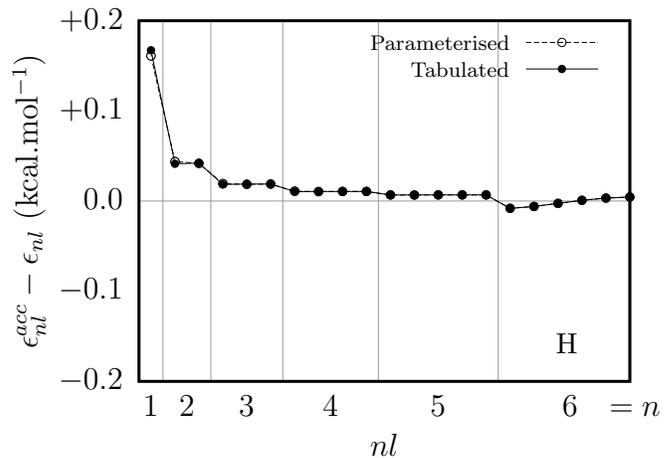}
\caption{ \label{fig2} Difference between `accurate' and CEPP 
  excitation energies for the single-electron H atom.  Both tabulated 
  and parameterized CEPP results are shown, and the `accurate' data 
  are experimental data taken from NIST\cite{NIST_De}.
  Results are shown in the range $n=1\ldots6$ and $l=0\ldots5$, with 
  each segment containing results ordered by increasing $l$.
  }
\end{figure}

Figure~\ref{fig2} shows results for the neutral H atom\cite{NIST_H_I}
with the tabulated and parameterized CEPP.  All of the results are
well within chemical accuracy of $1$ kcal$.$mol$^{-1}$, and the
results obtained with the parameterized and tabulated CEPPs are almost
indistinguishable, with the maximum difference being $-0.006$
kcal$.$mol$^{-1}$ for $nl=1s$.  
Figure~\ref{fig2} does not show perfect agreement between the
`accurate' and CEPP excitation energies, even for the states used to
construct the CEPP, with the largest difference between the `accurate'
and pseudo-levels being $0.17$ kcal$.$mol$^{-1}$ at $nl=1s$.
This difference arises almost entirely from our assumption of an
infinite nuclear mass and, on adjusting the energy levels using the
appropriate scaling factor, the maximum value of the remaining error
is $-0.013$ kcal$.$mol$^{-1}$ for $nl=6s$.

\begin{figure}[t]
\includegraphics[scale=1.00]{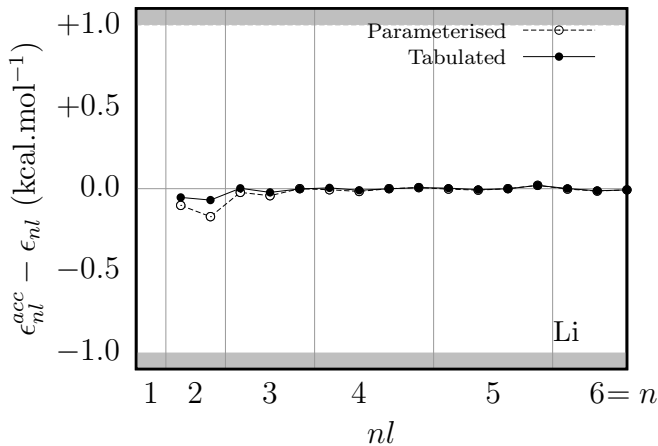}
\caption{ \label{fig3} Difference between `accurate' and CEPP
  excitation energies for the single-valence-electron Li atom.  Both
  tabulated and parameterized CEPP results are shown.  The `accurate'
  data are experimental\cite{NIST_De}.  Results are shown in the range
  $n=2\ldots6$ and $l=0\ldots5$ where `accurate' data are available.
  Each segment contains results ordered by increasing $l$.  }
\end{figure}

Figure~\ref{fig3} shows results for the neutral Li atom as the
difference between the `accurate' energy levels of Li\cite{NIST_Li_I}
and pseudo-levels obtained with the tabulated and parameterized CEPPs.
The `accurate' results are reproduced to well within chemical
accuracy, with the largest error for the tabulated and parameterized
CEPPs being $0.070$ kcal$.$mol$^{-1}$ (at $nl=2p$) and $0.17$
kcal$.$mol$^{-1}$ (at $nl=2p$), respectively.
The results obtained with the parameterized and tabulated CEPPs agree
to well within chemical accuracy, with a maximum difference of $-0.10$
kcal$.$mol$^{-1}$ at $nl=2p$.

\begin{figure}[t]
\includegraphics[scale=1.00]{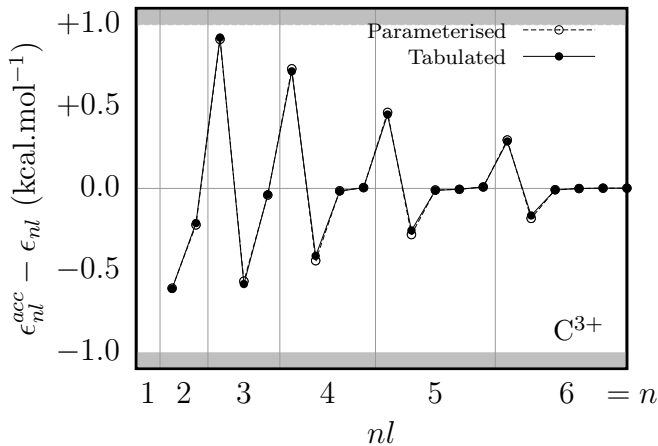}
\caption{ \label{fig4} Same as Fig.~\ref{fig3}, but for C$^{3+}$.}
\end{figure}
   
Figure~\ref{fig4} shows analogous results for the C$^{3+}$ ion.  The
difference between the `accurate'\cite{NIST_C_IV} and pseudo-atom
energy levels is larger than for Li, reflecting a general trend that
the error increases with atomic number (for the atoms considered).
The largest deviations from the `accurate' data for tabulated and
parameterized CEPPs are both  within chemical accuracy at $-0.89$
kcal$.$mol$^{-1}$ (at $nl=3s$).  The maximum difference between the 
tabulated and parameterized CEPP results of $-0.029$ kcal$.$mol$^{-1}$ 
for $nl=4p$ is smaller than the maximum difference for Li.

\begin{figure}[t]
\includegraphics[scale=1.00]{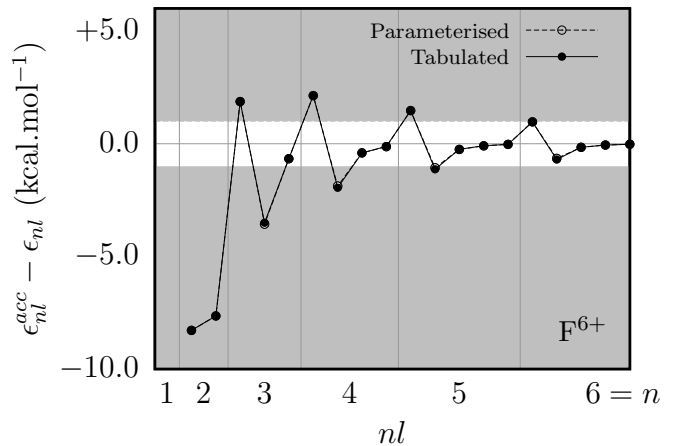}
\caption{ \label{fig5} Same as Fig.~\ref{fig4}, but for F$^{6+}$.
}
\end{figure}

Figure~\ref{fig5} shows similar results for the F$^{6+}$ ion.
The parameterization is accurate, with a maximum difference between 
the tabulated and parameterized CEPP results of $-0.074$ 
kcal$.$mol$^{-1}$ for $nl=4p$, which is smaller than that for Li.
The difference between the `accurate'\cite{NIST_F_VII} and 
pseudo-atom energy is larger than chemical accuracy for a number 
of levels, with differences of $-8.3$, $-7.7$, and $-0.7$ 
kcal$.$mol$^{-1}$ for $nl=2s,2p$ and $3d$.  This is not an error 
due to the CEPP generation process \emph{per se}, since the CEPP 
is designed to reproduce the properties of an ion with the core 
of the neutral atom.
For example, consider the largest error at $nl=2s$.
Generating an alternative CEPP from fully relaxed core NOs reduces 
this from $-7.7$ to $-4.5$ kcal$.$mol$^{-1}$, in good 
agreement with the AE MCHF value of $-4.6$ kcal$.$mol$^{-1}$ 
(and the AE CCSD(T) value of $-4.5$ kcal$.$mol$^{-1}$).
Furthermore, this remaining error is almost entirely due to 
relativistic effects; including a Breit-Pauli relativistic 
correction in the AE MCHF calculation results in a final 
error of only $-0.1$ kcal$.$mol$^{-1}$ (AE CCSD(T) with a 
Douglas-Kroll-Hess Hamiltonian results in an error of $0.2$ 
kcal$.$mol$^{-1}$).
This suggests that the deviation of the CEPP results from `accurate' 
results is satisfactory, since our CEPPs have been generated 
to represent the cores of the neutral atoms, and to exclude 
relativistic effects.

We conclude that our parameterization is successful, in that the
deviations of the pseudo-levels from `accurate' data are 
sufficiently small, and the CEPPs accurately describe isolated 
single-valence-electron atoms.
We also conclude that the deviation of the CEPP single-valence 
electron excitations from `accurate' results is dominated by 
physical effects absent from the MCHF data and, for fluorine, by 
the fixed core correction, rather than due to deficiencies in 
the pseudopotential generation procedure.
Note that these results provide no information on the transferability 
of the CEPPs between the ionic states and more neutral states.

\subsection{Ionization energies and electron affinities \label{ion_aff}}

Here we examine the accuracy of the parameterized CEPPs with more than
one electron in accounting for the ionization energies and electron
affinities of isolated atoms and ions.
Unlike the one-electron spectra considered above, we primarily compare
with AE results, although `accurate' data obtained from the NIST
online database\cite{NIST_De} are also considered.

Results were obtained at the CCSD(T) level using the aug-cc-pVnZ basis
set for H, and the aug-cc-pwCVnZ basis set otherwise, with the
complete basis set limit energies estimated using $n=T,Q$\cite{basis}
and Eqs.\ (\ref{extrap}).
Results were generated for AE atoms, and for the TNDF and BFD
potentials and the CEPPs.
Calculated ionization energies and electron affinities, together with
`accurate' data\cite{NIST_De}, are shown in Tables~\ref{tab:1}
and \ref{tab:2} for H, Li, C$^{3+}$, and F$^{6+}$. In what follows
we concentrate on the differences between the CEPP and AE results.

The data for H shown in Table \ref{tab:1} provide ionization energies
and electron affinities well within chemical accuracy of the AE
values, with the CEPP results deviating from the AE data by less than
$0.0035$ eV, and consistently being the most accurate of the three
pseudopotentials.
As for single-electron energy levels, the assumption of infinite nuclear 
mass results in an overestimate of the ionization energy by about 
$0.0073$ eV when compared with the `accurate' value, see Table 
\ref{tab:1}.
On adjustment for finite nuclear mass, the ionization energy arising 
from the CEPP calculation differs from the `accurate' result by 
only $0.004$ eV.

\begin{table}[b]
\begin{tabular}{lll}                                         \\ \hline \hline
\multicolumn{3}{c}{ H }                                      \\
  & \multicolumn{1}{c}{ $I_1$ (eV) } & \multicolumn{1}{c}{ EA (eV) }   \\ \hline
`Accurate'      & $13.598433770784(12)$ &  $0.754195(19)$    \\
AE              & $13.6066(3)$          &  $0.7471(8)$       \\
TNDF            & $13.6104(6)$          &  $0.7475(8)$       \\
BFD             & $13.6102(6)$          &  $0.7495(8)$       \\
CEPP            & $13.6098(6)$          &  $0.7469(8)$       \\ \hline \hline
\multicolumn{3}{c}{ Li }                                     \\
  & \multicolumn{1}{c}{ $I_1$ (eV) } & \multicolumn{1}{c}{ EA (eV) } \\ \hline
`Accurate'      & $5.391714668(22)$     & $0.618049(2)$      \\
AE              & $5.3921(2)$           & $0.6178(1)$        \\
TNDF            & $5.34289(2)$          & $0.62685(8)$       \\
BFD             & $5.34244(2)$          & $0.62719(8)$       \\
CEPP            & $5.38746(2)$          & $0.62555(9)$       \\ \hline \hline
\end{tabular} \ \ 
\caption{ \label{tab:1} 
  Ionization energies ($I$) and electron affinities (EA) of H and Li.
  All-electron and pseudopotential ionization energies and electron
  affinities are calculated with RCCSD(T).
  `Accurate' ionization energies are taken from NIST\cite{NIST_De}, 
  and are \emph{ab initio} for H and experimental for Li.
  The `accurate' electron affinity is from 
  experiment\cite{aff_H,aff_Li}.}
\end{table}

As demonstrated by the data in Table \ref{tab:1}, the Li CEPP performs
very well, with the deviation from the AE results being well within
chemical accuracy, with a maximum value of $0.008$ eV.  The agreement
with the AE results is much worse for the TNDF and BFD
pseudopotentials, with the first ionization energy deviating from the
AE value by about $0.043$ eV.  We conclude that the Li CEPP gives
more accurate results than the TNDF and BFD pseudopotentials.

The errors for C$^{3+}$ deduced from the data in Table \ref{tab:2} are
larger and more complex.
The largest deviation from the AE results occurs for the TNDF
pseudopotential ($I_4$), with the BFD pseudopotential providing the
smallest maximum deviation ($I_2$).
The CEPP and AE results agree to within chemical accuracy for three
out of the five cases, two of which are the important lowest energy
excitations of the first ionization energy and the electron affinity.
Neither the BFD or TNDF results for the first ionization energy agree
with the AE data to within chemical accuracy.

Table \ref{tab:2} also provides data for F$^{6+}$, showing similar 
behavior to the C$^{3+}$ case.  Both the TNDF and BFD pseudopotentials 
reproduce the AE results to within chemical accuracy for only the 
electron affinity, whereas the CEPP reproduces both the first 
ionization energy and electron affinity to within chemical accuracy.

There are two separate sources of error in using a CEPP for isolated
atoms: $(i)$ the error due to representing core-valence correlation by
a static potential, and $(ii)$ the error due to generating the CEPP
from an ion and using it for states that are close to neutral.
To investigate the relative magnitudes of these errors we generate a
second CEPP from a coreless one-electron Li$^{2+}$ ion using
the $2s\ (2S)$, $2p\ (2P)$, and $3d\ (2D)$ configurations, so that the
$s$-channel reproduces the $2s$ orbital.
This choice is made because it is much more important to accurately
reproduce the chemically active $2s$ orbital than the $1s$ core
orbital.
We choose the same core radii as in earlier
work\cite{Trail_2005_pseudopotentials}; $r_c=0.5$ a.u.\ for the
$s$-channel and $r_c=0.8$ a.u.\ for the rest.
(Results obtained with the coreless Li$^{2+}$ CEPP are not included 
in Table \ref{tab:1}.)

The second and third ionization energies obtained with the coreless
Li$^{2+}$ CEPP are not accurate as they differ from the AE and
`accurate' values by about $0.35$ eV.  
However, the first ionization energy and electron affinity are
considerably more accurate than those from the Li CEPP, with an error
of less than $0.0009$ eV (in comparison to $0.008$ eV from the CEPP
with a core).
It is clear that the coreless Li$^{2+}$ CEPP does not represent 
core-valence interactions with a static potential (as there are no 
core electrons) and is generated from an ion, whereas the Li CEPP 
describes core-valence interaction with a static potential and is 
constructed from a neutral atom.  We conclude that, at least for 
Li, the error due to transferring the CEPP between different 
ionization states is less than that due to representing 
core-valence correlation by a static potential.
However, for both cases the error is small and is well within 
chemical accuracy.

The errors for carbon and fluorine include contributions from 
representing the core-valence interaction by a static potential and 
from generating the CEPP in a highly ionized state and applying it to 
less ionized states.  Our results suggest that the sum of these 
errors is small since both the C$^{+3}$ and F$^{+6}$ CEPPs reproduce 
the AE values of the electron affinity and first ionization energy to 
within chemical accuracy.  Finally we note that, for Li, C$^{+3}$, 
and F$^{+6}$, considering the deviation of the CEPP results from 
`accurate' data does not alter our analysis significantly, and the AE 
results are within chemical accuracy of the `accurate' data.

We conclude that CEPPs successfully reproduce ionization energies and 
electron affinities for isolated atoms with one or more valence 
electrons.

\begin{table*}[b]
\begin{tabular}{llllll}                                                                                \\ \hline \hline
\multicolumn{6}{c}{ C$^{3+}$ }                                                                         \\
  & \multicolumn{1}{c}{ $I_4$ (eV) } & \multicolumn{1}{c}{ $I_3$ (eV) }
  & \multicolumn{1}{c}{ $I_2$ (eV) } & \multicolumn{1}{c}{ $I_1$ (eV) }
  & \multicolumn{1}{c}{ EA (eV) }                                                                      \\ \hline
`Accurate'        & $64.49358(19)$ & $47.88778(12)$ & $24.3845(9)$   & $11.26030$    & $1.262$         \\
AE                & $64.4674(5)$   & $47.874(1)$    & $24.384(4)$    & $11.260(2)$   & $1.250(1)$      \\
TNDF              & $64.112058(5)$ & $47.8901(9)$   & $24.249(3)$    & $11.191(2)$   & $1.229(1)$      \\
BFD               & $64.45501(5)$  & $48.0007(7)$   & $24.225(3)$    & $11.180(2)$   & $1.221(1)$      \\
CEPP              & $64.46782(5)$  & $48.1702(8)$   & $24.479(3)$    & $11.276(2)$   & $1.226(1)$      \\ \hline \hline
\multicolumn{6}{c}{ F$^{6+}$ }                                                                         \\
  & \multicolumn{1}{c}{ $I_4$ (eV) } & \multicolumn{1}{c}{ $I_3$ (eV) }
  & \multicolumn{1}{c}{ $I_2$ (eV) } & \multicolumn{1}{c}{ $I_1$ (eV) }
  & \multicolumn{1}{c}{ EA (eV) }                                                                      \\ \hline
`Accurate'        & $87.175(17)$   & $62.7080(3)$   & $34.97081(12)$ & $17.42282(5)$ & $3.4011895(25)$ \\
AE                & $87.209(7)$    & $62.799(7)$    & $34.962(10)$   & $17.425(6)$   & $3.431(5)$      \\
TNDF              & $87.072(4)$    & $62.696(4)$    & $34.873(9)$    & $17.370(5)$   & $3.413(5)$      \\
BFD               & $-$            & $62.624(5)$    & $34.863(10)$   & $17.371(6)$   & $3.411(5)$      \\
CEPP              & $87.528(4)$    & $63.007(4)$    & $35.033(9)$    & $17.436(5)$   & $3.412(5)$      \\ \hline \hline
\end{tabular}
\caption{ \label{tab:2}
  Ionization energies ($I$) and electron affinities (EA) of C and F.
  The all-electron and pseudopotential ionization energies and 
  electron affinities are calculated using RCCSD(T).  `Accurate' 
  ionization energies are taken from NIST\cite{NIST_De}, and are 
  experimental, with the exception of C($I_2$), F($I_4$), and F($I_3$), 
  which are semi-empirical.  The `accurate' electron affinity is from 
  experiment\cite{aff_C,aff_F}. }
\end{table*}

\subsection{Optimized geometries \label{geometry}}

Molecular geometries are obtained by direct minimization of the
CCSD(T) energy, enforcing the known symmetry of each molecule to
obtain the free parameters for optimization.
We generate results for the test set of $35$ small molecules for the
AE systems, and the TNDF and BFD pseudopotentials, and the CEPPs.

We have not attempted to extrapolate the geometries to the complete
basis set limit, and we use the basis sets as described following 
Eqs.\ (\ref{extrap}) with $n=Q$, except for H$_2$CO, H$_2$O$_2$, 
H$_3$COH, and H$_4$N$_2$, for which we use $n=T$.
The molecular geometries are characterized by their bond lengths, bond
angles, and dihedral angles.
Deviations of these quantities from the AE CCSD(T) values are evaluated, 
and we seek standard chemical accuracy of $0.01$ \AA.
Comparison of bond lengths is straightforward, but comparison of bond
angles and dihedral angles is less so.
We have mapped the bond and dihedral angles to the arc of a circle of
radius $1.0$ \AA\ (a typical bond length for the molecules
considered), and this arc length is compared to the standard chemical
accuracy value of $0.01$ \AA. To put this another way, we consider
chemical accuracy for bond and dihedral angles to be $0.57^\circ$.

\begin{figure*}[t]
\includegraphics[scale=1.00]{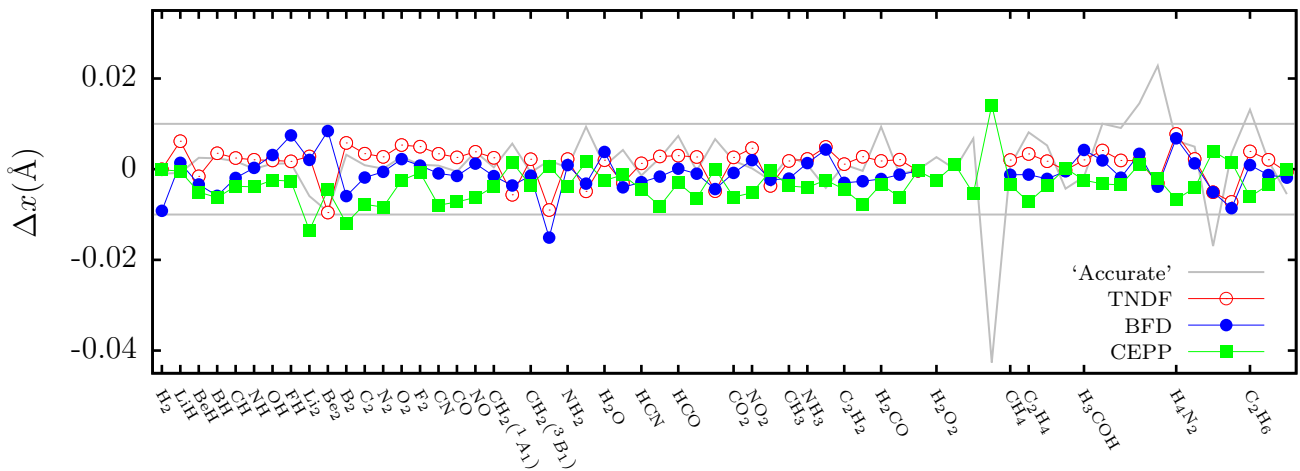}
\caption{ \label{fig6} Deviation of spatial parameters from the
  baseline AE CCSD(T) results.  The AE and pseudopotential results 
  are obtained by geometry optimization as described in the text, 
  using the TNDF and BFD pseudopotentials, and the CEPPs.
  The deviation of the `accurate' results from AE data is shown in 
  gray, with the `accurate' data taken from NIST\cite{NIST_geom}, 
  with the exception of CH$_2 (^3B_1)$\cite{CH2-2_geom}, 
  H$_2$O\cite{H2O_geom}, HCO\cite{HCO_geom}, and 
  H$_2$O$_2$\cite{H2O2_geom}.
  All `accurate' spatial parameters are experimental, and the 
  coordinates for each molecule are ordered as in the NIST database.
  Almost all of the pseudopotential results are within chemical 
  accuracy (horizontal gray lines) of the AE data.  }
\end{figure*}

Figure \ref{fig6} shows geometry parameters from optimizations with
all three pseudopotentials, as deviations from the baseline AE
results.
(No data are available for H$_2$O$_2$ with TNDF or BFD since these
calculations failed to converge.)
For almost all molecules the TNDF and BFD potentials, and the CEPPs,
provide molecular geometries within chemical accuracy of the AE results.
The maximum deviations from the AE results for the TNDF and BFD
potentials, and the CEPPs, are $0.0096$ \AA\ (for Be$_2$),
$0.015$ \AA\ (for CH$_2 (^3B_1)$), and $0.014$ \AA\ (for H$_2$O$_2$),
respectively.
The mean absolute deviations (MAD) from the AE results are 
$0.0032$ \AA, $0.0029$ \AA, and $0.0041$ \AA , respectively.

It is desirable to find some measure of the contribution of
core-valence interaction and transferability errors to the deviation
of the CEPP optimized geometries from the AE values.
For LiH and Li$_2$ this can be achieved by briefly returning to the
coreless one-electron Li$^{2+}$ CEPP of Sec.\ \ref{ion_aff}.
For LiH the deviation of the optimum bond length from the AE value is
reduced by $\times 1/6$ when we replace the `standard' Li CEPP by
the coreless Li$^{2+}$ CEPP.
Similarly, for Li$_2$ the coreless Li$^{2+}$ CEPP reduces this error
by $\times 1/9$.
Following the same reasoning used for the ionization energies, this
suggests that the error from the CEPP is mostly due to representing
core-valence interaction with a static potential, and that generating
a CEPP from an ion and transferring it to a neutral system introduces
a relatively small error.  (Results obtained using the coreless
Li$^{2+}$ CEPP are not shown in Fig.\ \ref{fig6}.)

We conclude that geometry optimization with the TNDF and BFD
potentials and the CEPPs is successful in that they reproduce the AE
results to within $0.015$ \AA.
The TNDF potentials give slightly more accurate geometries than the
BFD and CEPP potentials, and the TNDF geometries are within chemical
accuracy of the AE results for all of the systems studied.
However, all three potentials give a good description of the
geometries, with the variation in the errors between molecules being
comparable to the variation in the errors between pseudopotential
types.

\subsection{Well depths \label{atomization}}

The molecular well depth, $D_e$, is obtained at
the optimum geometry by evaluating the difference between the CCSD(T)
total energies of the molecules and their component atoms using
consistent basis sets, and estimating the energies in the complete
basis set limit using Eqs.\ (\ref{extrap}).
The basis sets used are as described following Eqs.\ (\ref{extrap}).

\begin{figure*}[t]
\includegraphics[scale=1.00]{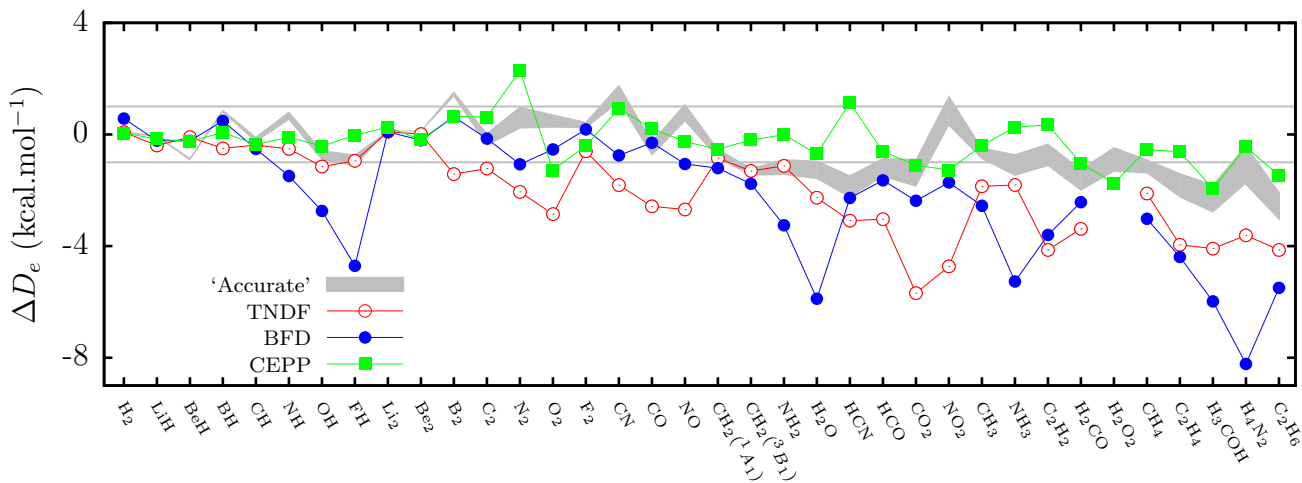}
\caption{ \label{fig7} Deviation of well depths from the baseline AE
  CCSD(T) results.  The AE and pseudopotential results are obtained at
  the optimum geometries as described in the text, using the TNDF and
  BFD pseudopotentials, and the CEPPs.  `Accurate'
  well depths are obtained by summing experimental dissociation
  energies and ZPVEs, with the exception of C$_2$\cite{Feller_2010}
  for which the well depth is \emph{ab initio}\cite{Feller_2010}.
  Dissociation energies are from O'Neil and Gill \cite{oneil_De},
  except for BH\cite{BH_De}, Be$_2$\cite{Be2_De}, B$_2$\cite{B2_De},
  and NO$_2$\cite{NO2_De}.  Zero-point vibrational energies are from
  NIST\cite{NIST_geom} except for Be$_2$\cite{Be2_De},
  B$_2$\cite{B2_De}, and NO\cite{NO2_De}.  The gray range shows the
  deviation of the `accurate' well depths from the range of AE well
  depths given by Eqs.\ (\ref{extrap}), and the gray horizontal lines
  indicate chemical accuracy.}
\end{figure*}

Figure~\ref{fig7} shows well depths, $D_e$, for the TNDF and BFD
potentials and the CEPPs, as deviations from the AE well depths.
All of the data are calculated at the optimum geometries (no data is
available for H$_2$O$_2$ with the TNDF or BFD pseudopotentials, since
these calculations failed to converge).
Note that Eqs.\ (\ref{extrap}) provide a range of values for the error
in $D_e$, but this range is not discernable on the scale of the plot.
The smallness of this range is due to cancellation of extrapolation
errors between the AE and pseudopotentials results.

Overall, the errors in the TNDF and BFD results increase with
molecular size, with a maximum deviation from the AE results of $-5.7$
kcal$.$mol$^{-1}$ (for CO$_2$) and $-8.2$ kcal$.$mol$^{-1}$ (for
H$_4$N$_2$), respectively.
However, the MAD values are similar, with $2.0$ kcal$.$mol$^{-1}$ for
the TNDF pseudopotentials and $2.2$ kcal$.$mol$^{-1}$ for BFD.
The agreement with the AE results is, for both TNDF and BFD, well
outside of chemical accuracy for most molecules, and we ascribe this,
at least in part, to the absence of correlation in the generation of
these pseudopotentials.
Overall the TNDF pseudopotentials appear to be more accurate than BFD,
but not consistently so for all of the molecules considered.

Figure~\ref{fig7} also shows well depths, $D_e$, for CEPPs as
deviations from the baseline AE results.  
These results are consistently more accurate than for the uncorrelated
pseudopotentials, with a maximum deviation from the AE results of $2.3$
kcal$.$mol$^{-1}$ (for N$_2$) and a MAD of $0.6$ kcal$.$mol$^{-1}$.
The well depths of $27$ out of the $35$ molecules fall within chemical
accuracy of the AE values.

We quantify the error in $D_e$ due to core-valence interaction and
transferability in the same manner as for geometry optimization.
Replacing the `standard' Li CEPP by the coreless one-electron
Li$^{2+}$ CEPP of Sec.\ \ref{ion_aff} results in a $\times 1/30$ and
$\times 1/50$ reduction in the deviation of the CEPP results from the
AE results for LiH and Li$_2$, suggesting that the error in the CEPP 
arises mostly from the representation of the core-valence interaction 
by a potential and that the error due to transferring the CEPP from 
an ion to the neutral system is small.  
(Results obtained using the coreless Li$^{2+}$ CEPP are not shown in
Fig.\ \ref{fig7}.)

We conclude that the CEPPs provide significantly more accurate well
depths than the TNDF and BFD pseudopotentials.

\subsection{Zero-point vibrational energies \label{vibrational}}

The harmonic ZPVEs are obtained within CCSD(T) by diagonalization of
the Hessian obtained from numerical energy derivatives at the optimum
geometry, and summation of the contributions from each mode.
We do not attempt to extrapolate this data to the complete basis set
limit, and we use the basis sets as described following 
Eqs.\ (\ref{extrap}) with $n=Q$, except for B$_2$, N$_2$, H$_2$CO, 
H$_2$O$_2$, C$_2$H$_4$, H$_3$COH, H$_4$N$_2$, and C$_2$H$_6$, for 
which we use $n=T$.

\begin{figure*}[t]
\includegraphics[scale=1.00]{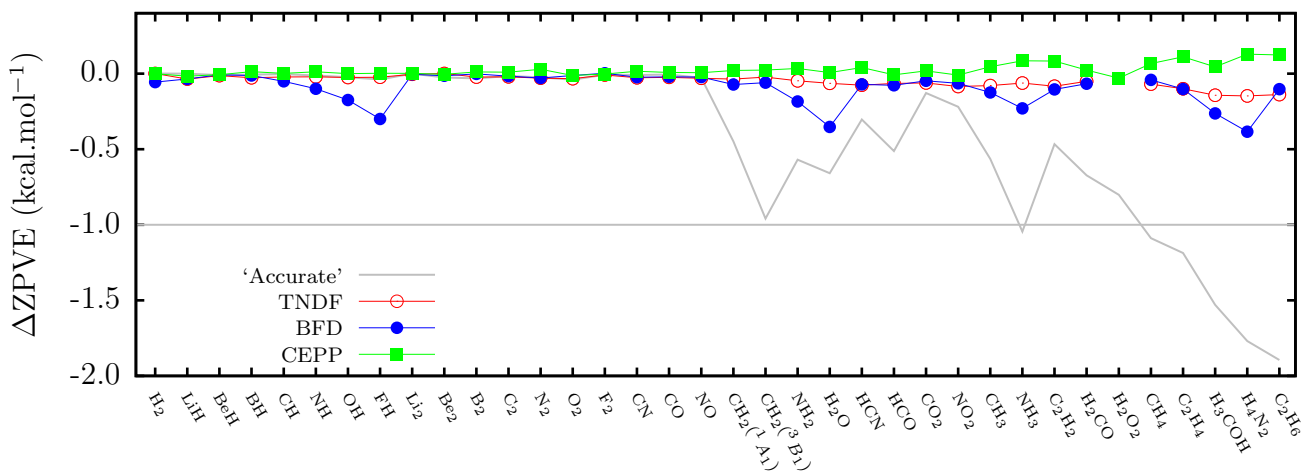}
\caption{ \label{fig8} Deviation of the ZPVEs from the baseline AE
  CCSD(T) results. The AE and pseudopotential results are obtained at
  the optimum geometries as described in the text, using the TNDF and
  BFD pseudopotentials and the CEPPs.  The deviation of the `accurate'
  experimental data from the AE results is shown in gray.  `Accurate'
  ZPVEs are from NIST\cite{NIST_geom}, except for Be$_2$\cite{Be2_De}.
  The gray horizontal line indicates the lower bound for chemical
  accuracy.  }
\end{figure*}

Figure~\ref{fig8} shows the ZPVEs obtained for the TNDF and BFD
potentials and the CEPPs as deviations from the baseline AE results
(no data is available for H$_2$O$_2$ with TNDF or BFD
pseudopotentials, as these calculations failed to converge).
For all three pseudopotentials and all molecules the ZPVEs
fall well within chemical accuracy (of $1$ kcal$.$mol$^{-1}$)
of the AE results.
Of the three pseudopotential types, the CEPPs consistently provide the 
most accurate ZPVEs.  The maximum deviations from the AE results for 
the TNDF and BFD potentials, and the CEPPs are
$-0.15$, $-0.38$, and $0.13$ kcal$.$mol$^{-1}$, respectively.
The MADs from the AE results are
$0.05$, $0.09$, and $0.03$ kcal$.$mol$^{-1}$, respectively.
It appears that the underestimation of the ZPVEs by the BFD
pseudopotentials for some molecules is primarily due to an inadequate
description of H.

We quantify the errors in the harmonic ZPVEs for the CEPPs due to 
core-valence interaction and transferability in the same manner as 
for the geometry optimization and well depth.
Replacing the `standard' Li CEPP by the coreless one-electron
Li$^{2+}$ CEPP of Sec.\ \ref{ion_aff} results in a $\times 1/143$ and
$\times 1/6$ reduction in the differences between the CEPP and AE
results for the LiH and Li$_2$ molecules.
This suggests that the error in the CEPP ZPVE is mostly due to the
representation of the core-valence interaction by a potential, with 
the error due to transferring the CEPP from an ion to a neutral 
system being relatively unimportant.
(Results obtained using the coreless Li$^{2+}$ CEPP are not shown in
Fig.\ \ref{fig8}.)

We conclude that, of the three pseudopotentials, the CEPPs reproduce
the AE ZPVEs to the highest accuracy.
The ZPVEs calculated using the TNDF pseudopotentials are marginally 
less accurate.  The ZPVEs calculated using the BFD pseudopotentials 
are significantly less accurate, with a maximum deviation from the AE 
results of $\times 3$ greater than for the CEPPs, but still well 
within chemical accuracy of the AE results.

\subsection{All-electron CCSD(T) and `accurate' data}
The differences between the `accurate' geometries and the AE data is
significantly larger than chemical accuracy for H$_2$O$_2$, H$_3$COH,
and H$_4$N$_2$, where the errors occur for bond or dihedral angles.
Relaxing the geometry of H$_2$O$_2$ using the contracted aug-cc-pVTZ
basis results in a negligible change in bond lengths, a small
improvement in the bond angle, and an error in the dihedral angle
which is smaller than chemical accuracy.
This suggests that the error in the original calculation arose from
the absence of diffuse basis functions resulting in a poor description
of bond angles involving H-H interactions.

Geometry optimization for H$_3$COH using the contracted aug-cc-pVTZ 
basis set suggests a similar source of error, resulting in a 
negligible change in bond lengths and an error in bond angles of 
chemical accuracy or less.
Geometry optimization of H$_4$N$_2$ using the contracted aug-cc-pVTZ
appears to increase the errors in both bond angles.
However, this is probably not significant given the uncertainty in 
the experimental values for these quantities\cite{H4N2_exp}.

There is an underlying trend for the well depths, $D_e$, to be
overestimated in the AE results as compared with the `accurate' data.
It seems reasonable to ascribe part of this error to the absence of
relativistic effects since the relativistic correction provided by
O'Neil and Gill\cite{oneil_De} shows a similar general behavior and
magnitude, decreasing the well depth by $0.0-1.1$ kcal$.$mol$^{-1}$.
However, such a correction does not explain all of the error
(particularly for B$_2$, CN, and NO$_2$).

For a detailed analysis of the remaining error due to 
extrapolation to the complete basis set limit and correlation 
missing from CCSD(T) we refer the reader to Feller and 
Peterson\cite{feller_99}.

Our AE results agree with `accurate' data with a similar
accuracy to that achieved by Feller and Peterson, with a MAD and
maximum deviation from the `accurate' data of $0.95$ and $2.5$
kcal$.$mol$^{-1}$ for those molecules common to both their paper and 
ours (all of our set except for BH, Be$_2$, B$_2$, C$_2$, and 
NO$_2$).  For comparison, taking the results of Feller 
and Peterson that include core-valence correlation and comparing with 
our `accurate' data results in a MAD and maximum deviation of $0.53$ 
and $2.9$ kcal$.$mol$^{-1}$.
We consider the agreement between our \emph{ab initio} AE results
and `accurate' data to be as good as we could hope for, given 
the neglect of relativistic corrections, the correlation missing from 
CCSD(T) in the complete basis set limit (estimated\cite{feller_99} to 
be roughly $1$ kcal$.$mol$^{-1}$), and that the experimental 
errors\cite{feller_99}, when available, fall within the range 
$0.01-2.4$ kcal$.$mol$^{-1}$.

Overall the deviation of the `accurate' well depths from the AE
results is not significantly different from the deviation of the CEPP
well depths from the AE data; the MADs of the AE results from the
`accurate' data, the CEPP results from the `accurate' data, and the
CEPP results from the AE results are $1.0$, $0.8$, and $0.6$ 
kcal$.$mol$^{-1}$, respectively.

The MAD of the `accurate' ZPVEs from the AE results is $0.42$
kcal$.$mol$^{-1}$, with a maximum value of $-1.89$ kcal$.$mol$^{-1}$
for C$_2$H$_6$.
This is significantly larger than the deviation of the CEPP results
from the AE values ($0.03$ kcal$.$mol$^{-1}$, and a maximum of $0.13$
kcal$.$mol$^{-1}$ for H$_4$N$_2$).

The data shows the general trend that the calculated harmonic ZPVEs 
overestimate the `accurate' data, with the overestimation increasing 
with the number of H atoms present in each molecule.
This trend is particularly apparent for the 8 larger molecules on 
the right hand side of Fig.~\ref{fig8}.
It is also evident in the good agreement between the `accurate' data 
and the AE results for the two largest molecules containing no 
hydrogen atoms, CO$_2$ and NO$_2$, and in the small error for the 
diatomic molecules compared with the rest of the set.
Overall the deviation of the `accurate' ZPVEs from the AE data 
appears to be dominated by cubic anharmonic effects involving hydrogen 
atoms\cite{pfeiffer_13}, which reduce the vibrational frequencies.
Such effects are not included in our AE and pseudopotential 
calculations, suggesting that the AE harmonic ZPVEs have provided the 
appropriate baseline for assessing the performance of the 
pseudopotentials.

\section{Conclusions}

We have developed a scheme for generating pseudopotentials suitable
for use in correlated-electron calculations.  
These correlated electron pseudopotentials (CEPPs) are created using
data from correlated-electron atomic MCHF calculations and \emph{ab
  initio} core polarizabilities.
We have created CEPPs for the H, Li, Be, B, C, N, O, and F atoms,
although our approach can readily be applied to heavier elements.  We
emphasize that the full accuracy of the CEPPs is obtained only when
the potential of Eq.\ \ref{hamilcpp} is used,
made up of an \emph{ab initio} one-electron term and 
three many-body terms taken as part of the semi-empirical CPP potential.

The CEPPs have been tested by performing CCSD(T) calculations with
large Gaussian basis sets for various atoms and molecules and 
comparing the resulting equilibrium geometries, well depths, 
and zero-point vibrational energies of 35 small molecules with 
accurate AE results.
The MAD and maximum errors in the well depths of the 35
molecules are: CEPP ($0.6$ and $2.3$ kcal$.$mol$^{-1}$), TNDF ($2.0$
and $-5.7$ kcal$.$mol$^{-1}$), and BFD ($2.2$ and $-8.2$
kcal$.$mol$^{-1}$).
These results demonstrate the superior performance of our CEPPs for
correlated systems, as compared with the uncorrelated pseudopotentials 
available in the literature.  
The results for the geometries and ZPVEs are similar for the different
potentials.

Many of the CEPPs are generated in highly ionized states, but our
results show that they can give highly accurate results for neutral
systems.
In the light of the known transferability problems that occur for 
norm-conserving DFT pseudopotentials this is, perhaps, a 
surprising feature of our results.
This can be understood by noting that our CEPPs are constructed by exact 
inversion of the pseudo-atom Schr\"odinger equation, whereas a 
norm-conserving DFT pseudopotential is constructed by an approximate 
inversion of the self-consistent Kohn-Sham equations (in the sense that 
the exchange-correlation functional is approximate, contains 
self-interaction, and is linearized in the inversion process).
Consequently the CEPPs can be expected to show better transferability 
than DFT pseudopotentials.
Furthermore, it is well known that atomic cores become less responsive to 
valence electrons as we move to the right of each period (as in CPP theory).
This suggests that although the description of core-valence interactions 
will become less accurate it will also exhibit a weaker dependence on the 
behaviour of the valence electrons.
Our results are consistent with this; the transfer error for lithium was 
found to be negligible and as we move to the right of the period no 
consistent increase in error is apparent for electron affinities or first 
ionization energies of atoms, or for the geometries, well depths, or ZPVEs 
of molecules.

It would be possible to improve the CEPPs by including, for example,
relativistic effects, although they are small for the light atoms
considered here.  
Overall we conclude that the CEPPs work very well in the cases
considered and that they produce better results in correlated-electron
calculations than HF-based pseudopotentials available in the
literature.

Tabulated and parameterized forms of the CEPPs described in this paper
are given in the supplementary material.\cite{Supplemental}

\begin{acknowledgments}
  The authors were supported by the Engineering and Physical Sciences
  Research Council (EPSRC) of the UK.
\end{acknowledgments}

\appendix*
\section{Properties of reduced density matrices due to modified core NOs}

In Sec.\ \ref{theory_pseudo_atoms} a many-body wave function made up 
of valence determinants and modified core determinants was defined and 
used to construct a $p$-body density matrix.  This $p$-body density 
matrix was then reduced to a $n$-body density matrix in order to 
define the CEPP.

It is tempting to assume that since the core determinants of 
Sec.\ \ref{theory_pseudo_atoms} are zero outside of the core region 
the modified core determinants will not contribute to the 
$n$-body reduced density matrix (and CEPP) in this region.
This is not so.  To demonstrate this we consider the special case 
of a Hartree-Fock wave function, which is a single normalized core 
determinant, for which the core NOs are chosen to be zero outside 
of the core region.

We define the Hartree-Fock wave function as a function of all 
electronic co-ordinates, $\mathbf{r}_1 \ldots\mathbf{r}_p$, where 
$|\mathbf{r}_{i=1 \ldots n}| > r_c$, but $\mathbf{r}_{i=n+1 \ldots p}$ 
are free to vary over all space.  The Slater determinant may then be 
written as 
\begin{equation}
\Psi_{\rm HF} =
\frac{1}{\sqrt{p!}}
\left| \begin{array}{cc}
\mathbf{A}_{11} & \mathbf{A}_{12} \\
\mathbf{0}      & \mathbf{A}_{22}
\end{array} \right| ,
\end{equation}
where row and column indices are orbital and electron co-ordinate 
indices, respectively.  The block-matrices $\mathbf{A}_{11}$ and 
$\mathbf{A}_{12}$ are composed of non-core orbitals, the block matrix 
$\mathbf{A}_{22}$ is composed of core orbitals, and the zero-block 
arises from the constraint on the first $n$ co-ordinates and the 
properties of the core orbitals.  Note that $\mathbf{A}_{11}$ is a 
$n \times n$ matrix, $\mathbf{A}_{12}$ is $n \times (p-n)$, and 
$\mathbf{A}_{22}$ is $(p-n) \times (p-n)$.

From the properties of zero-block matrices such a determinant may be 
written as
\begin{equation}
\Psi_{\rm HF} = \frac{1}{\sqrt{p!}} | \mathbf{A}_{11} || \mathbf{A}_{22} | ,
\end{equation}
from which it follows that the $p$-body density matrix can be written 
as
\begin{eqnarray}
\Gamma^p &=& \Psi^*_{\rm HF}( \mathbf{r}_1 \ldots\mathbf{r}_p )
             \Psi_{\rm HF}( \mathbf{r}_1'\ldots\mathbf{r}_p') \nonumber \\
&=& \frac{1}{p!}
| \mathbf{A}_{11}(\mathbf{r}_1     \ldots\mathbf{r}_n)  |^*
| \mathbf{A}_{22}(\mathbf{r}_{n+1} \ldots\mathbf{r}_p)  |^* \nonumber \\ & &
| \mathbf{A}_{11}(\mathbf{r}_1'    \ldots\mathbf{r}_n') |
| \mathbf{A}_{22}(\mathbf{r}_{n+1}'\ldots\mathbf{r}_p') |.
\end{eqnarray}

We may then reduce this to a $n$-body density matrix by integration 
over the final $(p-n)$ co-ordinates,
\begin{eqnarray}
\Gamma^n &=& {p \choose n} 
             \int \Gamma^p d\mathbf{r}_{n+1}\ldots\mathbf{r}_p \nonumber \\
&=& \frac{1}{p!} {p \choose n}
| \mathbf{A}_{11}(\mathbf{r}_1     \ldots\mathbf{r}_n)  |^*
| \mathbf{A}_{11}(\mathbf{r}_1'    \ldots\mathbf{r}_n') | \\ & &
\int d\mathbf{r}_{n+1}\ldots\mathbf{r}_p
| \mathbf{A}_{22}(\mathbf{r}_{n+1} \ldots\mathbf{r}_p ) |^*
| \mathbf{A}_{22}(\mathbf{r}_{n+1} \ldots\mathbf{r}_p ) |,  \nonumber
\end{eqnarray}
and use orthonormality of orbitals to obtain
\begin{eqnarray}
\Gamma^n &=&
\frac{(p-n)!}{p!} {p \choose n}
| \mathbf{A}_{11}(\mathbf{r}_1     \ldots\mathbf{r}_n)  |^*
| \mathbf{A}_{11}(\mathbf{r}_1'    \ldots\mathbf{r}_n') |  \nonumber \\
 &=& \frac{1}{n!} | \mathbf{A}_{11}(\mathbf{r}_1     \ldots\mathbf{r}_n)  |^*
                  | \mathbf{A}_{11}(\mathbf{r}_1'    \ldots\mathbf{r}_n') | .
\end{eqnarray}
In this final expression $\Gamma^n$ is clearly the $n$-body density 
matrix associated with a Slater determinant of the $n$ non-core  
orbitals.  Note that all of the above equations and statements are 
correct only for $|\mathbf{r}_{i=1 \ldots n}| > r_c$, and for core NOs 
that are zero outside of the core region.

For a multi-determinant expansion, the expressions given above become 
more complicated, but the largest contribution to the final $n$-body 
density matrix is still provided by the determinant whose expansion 
coefficient has the largest absolute value.  No information provided 
by the non-core NOs is lost in the CEPP generation procedure.

\end{document}